\documentclass[acmsmall]{acmart}

\AtBeginDocument{%
  \providecommand\BibTeX{{%
    \normalfont B\kern-0.5em{\scshape i\kern-0.25em b}\kern-0.8em\TeX}}}

\setcopyright{acmcopyright}
\acmJournal{JACM}
\usepackage{multirow}
\usepackage{array}
\usepackage{setspace}
\usepackage{url}

\begin{document}

%%
%% The "title" command has an optional parameter,
%% allowing the author to define a "short title" to be used in page headers.
\title{The Future of Misinformation Detection: New Perspectives and Trends}

%%
%% The "author" command and its associated commands are used to define
%% the authors and their affiliations.
%% Of note is the shared affiliation of the first two authors, and the
%% "authornote" and "authornotemark" commands
%% used to denote shared contribution to the research.
\author{Bin Guo}
%\authornote{Both authors contributed equally to this research.}
\email{guob@nwpu.edu.cn}
%\orcid{1234-5678-9012}
%\authornotemark[1]
\affiliation{%
  \institution{Northwestern Polytechnical University}
  \city{Xi'an}
  \country{P.R.China}
}

\author{Yasan Ding}
\affiliation{%
  \institution{Northwestern Polytechnical University}
  \city{Xi'an}
  \country{P.R.China}}
%\email{dingyasan@163.com}

\author{Lina Yao}
\affiliation{%
  \institution{The University of New South Wales}
  \city{Sydney}
  \country{Australia}
}
%\email{}

\author{Yunji Liang}
\affiliation{%
 \institution{Northwestern Polytechnical University}
 \city{Xi'an}
 \country{P.R.China}}
%\email{liang_yunji@outlook.com}

\author{Zhiwen Yu}
\affiliation{%
  \institution{Northwestern Polytechnical University}
  \city{Xi'an}
  \country{P.R.China}}
%\email{zhiwenyu@nwpu.edu.cn}

%%
%% By default, the full list of authors will be used in the page
%% headers. Often, this list is too long, and will overlap
%% other information printed in the page headers. This command allows
%% the author to define a more concise list
%% of authors' names for this purpose.
\renewcommand{\shortauthors}{B.Guo, et al.}

%%
%% The abstract is a short summary of the work to be presented in the
%% article.
\begin{abstract}
The massive spread of misinformation in social networks has become a global risk, implicitly influencing public opinion and
 threatening social/political development. Misinformation detection (MID) has thus become a surging research topic in recent years. As a promising and rapid developing research field, we find that many efforts have been paid to new research problems and approaches of MID. Therefore, it is necessary to give a comprehensive review of the new research trends of MID. We first give a brief review of the literature history of MID, based on which we present several new research challenges and techniques of it, including early detection, detection by multimodal data fusion, and explanatory detection. We further investigate the extraction and usage of various crowd intelligence in MID, which paves a promising way to tackle MID challenges. Finally, we give our own views on the open issues and future research directions of MID, such as model adaptivity/generality to new events, embracing of novel machine learning models, explanatory detection models, and so on.
\end{abstract}

%%
%% The code below is generated by the tool at http://dl.acm.org/ccs.cfm.
%% Please copy and paste the code instead of the example below.
%%


%%
%% Keywords. The author(s) should pick words that accurately describe
%% the work being presented. Separate the keywords with commas.

%%
%% This command processes the author and affiliation and title
%% information and builds the first part of the formatted document.
\maketitle

\section{Introduction}
Social media platforms (such as Twitter\footnote{\url{https://twitter.com/}}, Facebook\footnote{\url{https://www.facebook.com/}}, Sina Weibo\footnote{\url{https://weibo.com/}} etc.) have revolutionized the dissemination mode of information, which greatly improve the velocity, volume, and variety of information transmission. However, while the social platform accelerates the disclosure of information, it also brings the proliferation of misinformation. According to a recent survey of Knight Foundation\footnote{\url{https://www.poynter.org/ethics-trust/2018/americans-believe-two-thirds-of-news-on-social-media-is-misinformation/}}, Americans estimate that 65\% of the news they see on social media is misinformation. Besides, misinformation usually spreads faster, deeper, and wider in social networks \cite{vosoughi2018spread}.

The adversarial use of social media to spread deceptive or misleading information poses a political threat\cite{bakdash2018future}. For example, during the 2016 US presidential election, as many as 529 different rumors were spreading on Twitter \cite{jin2017detection}, and approximately 19 million malicious bot accounts published or retweeted tweets supporting Trump or Clinton\footnote{\url{https://firstmonday.org/ojs/index.php/fm/article/view/7090/5653}}, which potentially influenced the election. In 2018, the \textit{Science} magazine has published a theme issue about `Fake News', where they report that fake stories can arouse people's feelings of fear and surprise \cite{vosoughi2018spread}, which will contribute to social panic. For instance, a fake video named \textit{Somalis `pushed into shallow grave' Ethiopia} caused violent clashes between two races in Ethiopia\footnote{\url{https://www.bbc.com/news/world-africa-46127868}} and a social media rumor which suggests that onward travel restrictions have been lifted in Greece resulted in the Greek police clash with migrants\footnote{\url{https://www.bbc.com/news/world-europe-47826607}}. The above examples show that the wide spread of fake news poses a serious threat to the ecology of social information dissemination \cite{lazer2018science}. Since infinite information and limited attention lead to the reduction of humans' ability to distinguish true information from misinformation \cite{qiu2017limited}, it is urgent to detect misinformation on social media.

The widespread misinformation not only misleads people to accept fake beliefs and changes the way they respond to the truth, but also breaks the trustworthiness of entire information ecosystem \cite{shu2018understanding}. What's worse, the future of online deception, including rumors, fake news and so on, will extend beyond text to high-quality and manipulative information materials, such as images, videos, and audios on a massive scale with the rapid development of artificial intelligence technology \cite{bakdash2018future}. For example, \textit{DeepFakes} \cite{guera2018deepfake, floridi2018artificial} utilizes deep learning algorithms to create audio and video of real people saying and doing things they never said or did, which makes misinformation ever more realistic and harder to discern. Therefore, while automatic misinformation detection is not a new phenomenon, it has been attracting much more public attention at present.

In recent years, there have been numerous efforts on misinformation detection (MID). According to the type of features used in existing MID methods, we divide them into four categories: content features, social context features, feature fusion, and deep learning-based methods. The content-based detection methods mainly utilize the textual or visual features extracted from the fake message for classification. The social context-based methods generally rely on the interaction characteristics among the majority of users, such as commenting, reposting, following etc. The feature fusion methods make comprehensive use of content features and social context features. Deep learning-based methods prevailingly learn the latent depth representation of misinformation through neural networks.

Though much effort has been made on MID in the past years, there are still numerous remaining issues to be addressed. First, existing MID methods mostly utilize content or propagation features and often work well on the entire life cycle of misinformation, which may contribute to poor performance for early detection. Since misinformation could have a severe impact in just a few minutes\footnote{\url{https://www.theverge.com/2013/4/23/4257392/ap-twitter-hacked-claims-explosions-white-house-president-injured}}, it is crucial for detecting them at the early stage. Second, with the increase of multimodal misinformation propagating on social networks, traditional text-based detection approaches are no longer practicable and it is beneficial to take advantage of images or videos for MID in more complex scenarios. Third, current detection methods only give the final result of whether the claim is false or not, but lack the reason for the decision. It is significant to give a convincing explanation for debunking misinformation and preventing its further propagation.

This paper aims to give an in-depth survey of the recent development of MID methods. There have been several surveys on MID \cite{zhou2018fake, zubiaga2018detection, shu2017fake}. Zhou \textit{et al.} \cite{zhou2018fake} study fake news from four perspectives, including knowledge-based, style-based, propagation-based and credibility-based, and summarize relevant detection methods in psychology and social science. Zubiaga \textit{et al.} \cite{zubiaga2018detection} focus on rumor classification systems and investigate the existing approaches for identifying suspected rumors, collecting rumor-related posts, detecting stances of posts, and evaluating the credibility of target events. Shu \textit{et al.} \cite{shu2017fake} divide detection models into news content-based models and social context-based models from the perspective of data mining, and summarize the evaluation measurements of fake news detection algorithms. The above surveys summarize MID studies from different perspectives, mostly aim to categorize the methods for solving the general MID problems. Differently, we intend to investigate this field from a problem-driven perspective, based on a brief review of the recent progress on solving general MID problem, we pay more attention to the new research trends of MID, characterized as the following three research problems, namely early detection, detection by multimodal data fusion, and explanatory detection. Furthermore, at the technical level, we particularly review a novel manner for MID --- crowd intelligence-based MID. Different from existing studies that mostly use the content of posts, crowd intelligence-based methods aim to detect misinformation based on aggregated user opinions, conjectures and evidences, which are implicit knowledge injected during human-post interaction (e.g., publishing, commenting, and reposting of posts).

Above all, the main contributions of our work include:
\begin{itemize}
\item Based on a brief literature review of MID, we concentrate on the recent research trends of it, including model generality to new events, early detection, multimodal fusion-based detection, and explanatory detection.
\item We make an investigation of the crowd-intelligence-based approach for MID, including the scope of crowd intelligence in MID, crowd-intelligence-based detection models, and hybrid human-machine fusion models.
\item We further discuss the open issues and promising research directions of MID, such as model adaptivity/generality to new events, embracing of novel machine learning models, and hybrid human-machine systems.
\end{itemize}

The rest of this paper is organized as follows. We give a brief literature review of existing MID work in Section 2. Then we investigate several new research trends in MID in Section 3. In Section 4, we highlight the crowd intelligence-based detection followed by open issues and future directions of MID in Section 5. Finally, we conclude this article in Section 6.

\section{A Brief Literature Review}
Various types of false information spread on social media platforms, such as fake news, rumor, hoax etc. Moreover, the definition of them varies over existing papers. The widely-recognized definitions are summarized in Table~\ref{tab:Definitions}.

\begin{table}
 \caption{Definitions of Some Types of False Information}
 \label{tab:Definitions}
 \begin{tabular}{|c|m{10.5cm}|}
    \hline
    \textbf{Term} & \textbf{Definition}\\
    \hline
    Rumor & ``\textit{An item of circulating information whose veracity status is yet to be verified at the time of posting}'' \cite{zubiaga2018detection}.\\
    \hline
    Fake News & ``\textit{A news article that is intentionally and verifiable false}'' \cite{shu2017fake}.\\
    \hline
    Hoax & ``\textit{A deliberately fabricated falsehood made to masquerade as truth}'' \cite{kumar2016disinformation}.\\
    \hline
    Click-bait & ``\textit{A piece of low-quality journalism which is intended to attract traffic and monetize via advertising revenue}'' \cite{volkova2017separating}.\\
    \hline
    Disinformation & ``\textit{Fake or inaccurate information which is intentionally false and deliberately spread}'' \cite{wu2016mining}. \\
    \hline
    Misinformation & ``\textit{Fake or inaccurate information which is unintentionally spread}'' \cite{wu2016mining}.\\
    \hline
  \end{tabular}
 \end{table}

In this work, we primarily focus on inaccurate information spreading on social networks which misleads people, so we define \textit{misinformation} as follows:

\begin{definition}
MISINFORMATION: \textit{Information or a story propagating through social media which is ultimately verified as false or inaccurate.}
\end{definition}

This definition covers different kinds of false information classified by their intention or motivation, such as profit, political intervention, deliberate deception and unintentional misrepresentation etc. We further give a general definition of the misinformation detection problem.
\begin{itemize}
\item For a specific story \textit{s}, it contains a set of related \textit{n} posts $ P=\{p_{1},p_{2},\cdots,p_{n}\} $ and a set of relevant \textit{m} users $ U=\{u_{1},u_{2},\cdots,u_{m}\} $. Each $ p_{i} $ consists of a series of attributes representing the post, including text, images, number of comments etc. Every $ u_{i} $ consists of a series of attributes describing the user, including name, register time, occupation etc.
\item Let $ E=\{e_{1},e_{2},\cdots,e_{n}\} $ refers to the engagements among \textit{m} users and \textit{n} posts. Each $ e_{i} $ is defined as $ e_{i} = \{p_{i},u_{j},a,t\} $ representing that a user $ u_{j} $ interacts with the post $ p_{i} $ through action $ a $ (posting, reposting, or commenting) at time $ t $.
\end{itemize}
\begin{definition}
MISINFORMATION DETECTION: \textit{Given a story $\textbf{s}$ with its posts set $\textbf{P}$, users set $\textbf{U}$ and engagements set $\textbf{E}$, the misinformation detection task is to learn a prediction function $ \mathcal{F}(\textbf{s}) \rightarrow \{0,1\} $, satisfying:}
$$
\mathcal{F}(\textbf{\textit{s}}) =
\begin{cases}
1, & \text{if $\textbf{\textit{s}}$ is a piece of misinformation} \\
0, & \text{otherwise}
\end{cases}
$$
\end{definition}

In the following, we give a brief literature review of existing MID techniques, categorized into four major types, namely \textbf{\textit{content-based}}, \textbf{\textit{social context-based}}, \textbf{\textit{feature fusion-based}}, and \textbf{\textit{deep learning-based}} methods, as summarized in Table~\ref{tab:Features} (for the prior three types) and Table~\ref{tab:Deep} (for the last type).

\begin{table}
 \caption{A Summary of Features Used by Existing Methods}
 \label{tab:Features}
 \begin{tabular}{|m{2.9cm}<{\centering}|m{4.8cm}<{\centering}|m{5cm}<{\centering}|}
 \hline
 \multirow{2}{*}{\textbf{Work}} & \multicolumn{2}{c|}{\textbf{Feature Type}} \\
 \cline{2-3}
  & \textbf{Content} & \textbf{Social Content} \\
 \hline
  Castillo \textit{et al.} \cite{castillo2011information} & \textit{Containing question marks, sentiment, URL links, etc.} & \textit{Propagation initial tweets, max subtree, average degree, etc.} \\
 \hline
  Gazvinian \textit{et al.} \cite{qazvinian2011rumor} & \textit{Unigram, bigram, trigram, POS, hashtag etc.} & \textit{Propagation structure} \\
 \hline
  Hu \textit{et al.} \cite{hu2014social} & \textit{Topic distribution, sentiment information} & \rule[0pt]{1cm}{0.05em} \\
 \hline
  Horne \textit{et al.} \cite{horne2017just} & \textit{Language complexity, stylistic features} & \rule[0pt]{1cm}{0.05em} \\
 \hline
  Gupta \textit{et al.} \cite{gupta2013faking} & \textit{First|second|third pronoun, exclamation marks, etc.} & \textit{Follower-friend ratio, number of friends} \\
 \hline
  Shu \textit{et al.} \cite{shu2018understanding} & \rule[0pt]{1cm}{0.05em} & \textit{Number of likes, number of followers, etc.} \\
 \hline
 \rule{0pt}{16pt}
  Tacchini \textit{et al.} \cite{tacchini2017some} & \rule[0pt]{1cm}{0.05em} & \textit{Number of likes} \\
 \hline
  Yang \textit{et al.} \cite{yang2019unsupervised} & \rule[0pt]{1cm}{0.05em} & \textit{User opinions, viewpoints, user credibility} \\
 \hline
  Ma \textit{et al.} \cite{ma2015detect} & \textit{Topic distribution, question marks, exclamation marks, etc.} & \textit{Average number of retweets and comments etc.} \\
 \hline
 \rule{0pt}{16pt}
  Liu \textit{et al.} \cite{liu2016detecting} & \rule[0pt]{1cm}{0.05em} & \textit{User credibility, friendship} \\
 \hline
 \rule{0pt}{16pt}
  Ma \textit{et al.} \cite{ma2017detect} & \rule[0pt]{1cm}{0.05em} & \textit{Syntactic parse tree and subtrees} \\
 \hline
  Jin \textit{et al.} \cite{jin2014news} & \textit{Hashtag topic, URL links, etc.} & \textit{Number of forwards and comments, propagation structure} \\
 \hline
 \rule{0pt}{16pt}
  Della \textit{et al.} \cite{della2018automatic} & \textit{TF-IDF, stemmer} & \textit{Number of likes} \\
 \hline
  Shu \textit{et al.} \cite{shu2017exploiting} & \textit{Content embedding} & \textit{News-user social engagement embedding} \\
 \hline
  Kwon \textit{et al.} \cite{kwon2013prominent} & \textit{Positive words, negating words, cognitive action words, inferring action words etc.} & \textit{Clustering of friendship network, fraction of isolated nodes etc.} \\
  \hline
  Wu \textit{et al.} \cite{wu2015false} & \rule[0pt]{1cm}{0.05em} & \textit{Number of followers, number of comments and reposts, propagation tree} \\
 \hline
  Volkova \textit{et al.} \cite{volkova2018misleading} & \textit{Language complexity and readability, moral foundations, psycholinguistic cues, etc.} & \textit{User opinions} \\
 \hline
 \end{tabular}
\end{table}

\subsection{Content-based Methods}
For a specific event, its microblog is generally composed of a piece of text to describe it, often associated with several pictures or videos. Content-based methods are mainly based on specific writing styles or sensational headlines in fake messages, such as lexical features, syntactic features and topic features \cite{rubin2015truth}. Compared with true information, the misinformation is fabricated to mislead the public and attract people's attention, so its content usually has disparate patterns. For example, Castillo \textit{et al.} \cite{castillo2011information, castillo2013predicting} find that highly credible tweets have more URLs and that the text length is usually longer than tweets with lower credibility.

Many studies utilize the lexical and syntactic features to identify misinformation. For instance, Qazvinian \textit{et al.} \cite{qazvinian2011rumor} find that the \textit{part of speech (POS)} is a distinguishable feature for rumor detection. Kwon \textit{et al.} \cite{kwon2013prominent} find that some types of sentiments are apparent features for rumor detection, including \textit{positive sentiments words} (e.g., love, nice, sweet), \textit{negating words} (e.g., no, not, never), \textit{cognitive action words} (e.g., cause, know) and \textit{inferring action words} (e.g., maybe, perhaps), and then they propose a periodic time series model to identify key linguistic differences between rumors and non-rumors.

Lexical features sometimes cannot fully reflect characteristics of misinformation because of its locality, and therefore many studies introduce novel features of topic, sentiment, and writing style for MID. For example, Potthast \textit{et al.} \cite{potthast2017stylometric} propose a meta-learning approach that utilizes different writing styles to detect fake news. Wu \textit{et al.} \cite{wu2015false} leverage LDA model to extract topic-level features from contents of Weibo for rumor detection. Hu \textit{et al.} \cite{hu2014social} propose a framework for social spammer detection with sentiment information. In \cite{horne2017just}, Horne \textit{et al.} propose a fake news detection model based on the observation that fake news is substantially different from real news in their title style.

\subsection{Social Context-based Methods}
Traditional content-based methods analyze the credibility of the single microblog or a piece of news in isolation, ignoring the high correlation of different tweets and events. However, lots of interactions among users or contents (adding friends, following, posting, commenting, reposting and tagging etc.) provide abundant reference information for the identification of misinformation. Concretely, these approaches are mainly divided into \textbf{\textit{post-based}} and \textbf{\textit{propagation-based}} methods.

(1) \textbf{\textit{Post-based features}}

Post-based methods mainly rely on users' posts which express their emotions or opinions related to the given event. Many studies detect misinformation by analyzing users' credibility \cite{morris2012tweeting} or stances \cite{mohammad2017stance, hanselowski2018retrospective}. For instance, Shu \textit{et al.} \cite{shu2018understanding} explore the truely useful features of user profiles for MID, so as to reduce the burden of feature extraction in the process of detection. Tacchini \textit{et al.} \cite{tacchini2017some} find that the social posts on Facebook can be detected as hoaxes according to the users' \textit{like} behaviors. Yang \textit{et al.} \cite{yang2019unsupervised} propose the fake news detection method called \textit{UFD}, which treats the authenticity of news and users' reputation as latent variables and exploits users' social engagements to extract their viewpoints on news credibility.

(2) \textbf{\textit{Propagation-based features}}

Propagation-based methods evaluate the credibility of messages and events as a whole \cite{cao2018automatic}, whose core is credibility network construction and credibility propagation.

Several existing studies detect misinformation by analyzing information propagation modes. For instance, Jin \textit{et al.} \cite{jin2013epidemiological} utilize the epidemiological models to describe the dissemination process of rumors on Twitter, and propose an enhanced epidemic model for MID. Ma \textit{et al.} \cite{ma2015detect} characterize the temporal patterns of social context features for fake news detection. Liu \textit{et al.} \cite{liu2016detecting} propose the information dissemination networks based on heterogeneous users' specific attributes for rumor detection. Kim \textit{et al.} \cite{kim2019homogeneity} propose a Bayesian nonparametric model to characterize the process of fake news transmission, jointly utilizing topics of news stories and user interests for MID.

There are also several studies that detect misinformation by constructing specific tree or network structures. For example, Ma \textit{et al.} \cite{ma2017detect} model the propagation of rumor-related microblogs as propagation trees, and they propose a kernel-based method to capture the patterns among those propagation trees for MID. In addition, Gupta \textit{et al.} \cite{gupta2012evaluating} construct a credibility propagation network containing users, messages and events for MID. In \cite{jin2014news}, Jin \textit{et al.} propose a three-layer credibility propagation network that connects microblogs, sub-events and events for news credibility validation.

\subsection{Feature Fusion-based Methods}
Content-based methods mainly classify the difference between true and false messages in terms of writing style, lexical and syntactic features, while social context-based methods mainly leverage the features extracted from information propagation for MID. Due to their complementary nature \cite{ruths2019misinformation}, recently there have been studies that try to combine them using feature fusion-based methods. For example, Vedova \textit{et al.} \cite{della2018automatic}  jointly leverage the ID of the users interacting with news and the textual information of the news item for the detection of fake news. In addition to utilizing traditional content features (e.g., lexical or syntactic features), Volkova \textit{et al.} \cite{volkova2018misleading} leverage psycho-linguistic signals from the content of fake news and authors' perspectives from social context as input data for different classifiers in MID. Shu \textit{et al.} \cite{shu2017exploiting} propose a semi-supervised multi-feature fusion-based model, which combines media bias, news attitude, and relevant user social engagements concurrently for fake news detection.

\subsection{Deep Learning-based Methods}
Deep learning-based methods aim to abstract deep representation of misinformation data automatically. At present, most work mainly utilize Recurrent Neural Networks (RNN) \cite{ma2016detecting} and Convolutional Neural Networks (CNN) \cite{yu2017convolutional} for MID.
\begin{table}
 \caption{A Summary of Deep Learning-Based Methods}
 \label{tab:Deep}
 \begin{tabular}{|m{2.8cm}<{\centering}|m{1.5cm}<{\centering}|m{1.6cm}<{\centering}|m{1.6cm}<{\centering}|m{1.8cm}<{\centering}|m{1.8cm}<{\centering}|}
 \hline
 \multirow{2}{*}{\textbf{Work}} & \multirow{2}{*}{\textbf{Model}} & \multicolumn{4}{c|}{\textbf{Data Inputs}} \\
 \cline{3-6}
  &  & \textbf{Text} & \textbf{Visual data} & \textbf{User response} & \textbf{User or website profiles} \\
 \hline
 \rule{0pt}{18pt}
  Ma \textit{et al.} \cite{ma2016detecting} & RNN & \checkmark &   & \checkmark &   \\
 \hline
 \rule{0pt}{18pt}
  Yu \textit{et al.} \cite{yu2017convolutional} & CNN & \checkmark &   & \checkmark &   \\
 \hline
 \rule{0pt}{18pt}
  Jin \textit{et al.} \cite{jin2017multimodal} & RNN & \checkmark & \checkmark & \checkmark &   \\
 \hline
 \rule{0pt}{18pt}
  Li \textit{et al.} \cite{li2018rumor} & GRU & \checkmark &   & \checkmark &   \\
 \hline
 \rule{0pt}{18pt}
  Liu \textit{et al.} \cite{liu2018mining} & Attention & \checkmark &   & \checkmark &   \\
 \hline
  Runchansky \textit{et al.} \cite{ruchansky2017csi} & RNN & \checkmark &   & \checkmark & \checkmark \\
 \hline
  Chen \textit{et al.} \cite{chen2018call} & LSTM + Attention & \checkmark &   & \checkmark &   \\
 \hline
  Nguyen \textit{et al.} \cite{nguyen2017early} & CNN + LSTM & \checkmark &   & \checkmark &   \\
 \hline
  Guo \textit{et al.} \cite{guo2018rumor} & LSTM + Attention & \checkmark &   & \checkmark & \checkmark \\
 \hline
  Popat \textit{et al.} \cite{popat2018declare} & LSTM + Attention & \checkmark &   &   & \checkmark \\
 \hline
  Liu \textit{et al.} \cite{liu2018early} & CNN + GRU &   &   & \checkmark &   \\
 \hline
 \end{tabular}
\end{table}

Many existing studies utilize deep neural networks to learn latent textual representation of misinformation by modeling related posts as time-series. For example, Ma \textit{et al.} \cite{ma2016detecting} propose a rumor detection model based on RNN which captures the temporal-linguistic features of a continuous stream of user comments related to each specific event. Li \textit{et al.} \cite{li2018rumor} consider that both the forward and backward sequences of post flow provide abundant interactive information, so they propose the Bidirectional Gated Recurrent Unit (GRU) method for rumor detection. Liu \textit{et al.} \cite{liu2018early} find that there are differences between the propagation patterns of fake news and true news, and they utilize CNN and GRU to classify the propagation paths for fake news detection.

There are also studies that combine textual information and social context information (such as user response, user or website profiles etc.) as data inputs of deep neural networks. For instance, Guo \textit{et al.} \cite{guo2018rumor} propose a hierarchical neural network that have information of users, posts, and the propagation network as data inputs, and attention mechanisms are leveraged to estimate the distinct contribution of them in MID. In \cite{ruchansky2017csi}, Ruchansky \textit{et al.} propose a MID model based on RNN, which incorporates features of the news content, the user response, and the source users to promote the performance on fake news detection.

\section{New Trends in Misinformation Detection}
Having reviewed the traditional studies on MID, this section investigates several new research trends of this field, including \textbf{\textit{early detection}}, \textbf{\textit{detection by multimodal data fusion}}, and \textbf{\textit{explanatory detection}}.

\subsection{Early Detection}
Misinformation can be readily spread by massive users on social networks, resulting in serious effects in a very short period \cite{cao2018automatic, friggeri2014rumor}. Therefore, early detection of misinformation becomes an important research topic. However, most existing studies debunk misinformation by assuming that they have all lifecycle content. They rely on the aggregation features, such as content characteristics and propagation patterns etc., which tend to emerge slowly and require a certain number of posts for classification. Therefore, the resource on the beginning of a piece of misinformation is so limited that it is challenging to detect it at the early stage. Recently, there have been some efforts for early detection of misinformation.

Many MID models leverage the deep learning techniques for early detection of misinformation. For example, Liu \textit{et al.} \cite{liu2018mining} observe that only a small number of posts contribute a lot to MID. In order to select these crucial contents, they propose an attention-based misinformation identification model, which evaluates the importance of posts by their attention values. Besides, the experiment results indicate that the usage of attention mechanism facilitates the early detection of misinformation. Similarly, Chen \textit{et al.} \cite{chen2018call} find that users tend to comment differently (e.g., from surprising to questioning) in different stages of the misinformation diffusion process. Based on this observation, a deep attention model based on RNN is proposed to learn selectively temporal hidden representations of sequential posts for early detection of rumors. Yu \textit{et al.} \cite{yu2017convolutional} utilize the CNN-based model to extract key features from the sequence of posts and learn high-level interactions among them, which is helpful for early detection of misinformation. Nguyen \textit{et al.} \cite{nguyen2017early} also leverage CNNs for learning the latent representations of each tweet, obtaining the credibility of individual tweets accordingly. They then evaluate whether the target event is a rumor by aggregating the predictions of related tweets at the very beginning of the event.

As for the lacking of data for early detection of misinformation, it will be a helpful method to borrow some knowledge from the related events. In \cite{sampson2016leveraging}, Sampson \textit{et al.} propose a method for emergent rumor detection by leveraging implicit linkages (e.g., hashtag linkages, web linkages) for employing additional information from related events. The experimental results show that such implicit links notably contribute to identifying emergent rumors correctly when a handful of textual or interactive data is available.

\subsection{Detection by Multimodal Data Fusion}
Traditional MID methods focus on textual content and propagation network. However, social media posts also contain rich media data such as images and videos, while such multimodal data are often ignored. Images and videos are more appealing to users compared to pure textual information because they can vividly describe target events. Analyzing the relationships among multimodal data and developing fusion-based models can be a promising way to MID \cite{cao2018automatic}. Beyond images and videos, it is also suggested to incorporate other auxiliary information to improve detection effect, such as user profiles, social contexts, etc. There have been recent studies about MID using multimodal data fusion.

Visual features have been used in different learning models for misinformation detection. For instance, Gupta \textit{et al.} \cite{gupta2013faking} analyze 10,350 tweets with fake images circulated during Hurricane Sandy in 2012, and find that the temporal characteristics, influence patterns and user responses in the process of image dissemination conduce to MID. Wang \textit{et al.} \cite{wang2018eann} and Jin \textit{et al.} \cite{jin2017multimodal} both propose the Recurrent Neural Network based detection models with an attention mechanism (att-RNN), which fuse the multimodal data (images, texts, and social contexts) and their interactions for rumor detection. In \cite{jin2016novel}, Jin \textit{et al.} observe that there are obvious differences in the image distribution patterns between real and fake news, and thus propose several visual and statistical features to characterize the differences for fake news detection. Sabir \textit{et al.} \cite{sabir2018deep} present a deep multimodal model for fake images detection, which simultaneously utilizes the convolutional neural network, word2vec, and globe positioning system to extract unique features of fake pictures.

Social context-based detection methods comprehensively utilize the characteristics from user profiles, post contents and social interactions for MID. For example, Shu \textit{et al.} \cite{shu2018understanding} delve into the relationship between user profiles and online fake news. The findings are useful for designing multimodal fusion MID systems. Then, Shu \textit{et al.} \cite{shu2019beyond} further explore the social relations among publishers, news pieces, and users, and propose a fake news detection framework called \textit{TriFN}, which models the inherent relationship among news content, social interactions, and news publishers by nonnegative matrix factorization (NMF) algorithms for MID.

The great advances in image processing techniques have proved that images can be easily edited and modified, making fake images generation more readily. For instance, the face of one person in videos can be automatically replaced with that of another without obvious sense of violation by using the pre-trained Variational Auto-Encoder (VAE) or Generative Adversarial Network (GAN). In \cite{korshunov2018deepfakes}, Korshunov \textit{et al.} investigate that existing \textit{Facenet}-based face recognition algorithms \cite{schroff2015facenet} are vulnerable to fake images and videos generated by GAN. Huh \textit{et al.} \cite{huh2018fighting} propose a fake image detection model based on image self-consistency features, which trains CNN using image exchangeable image file format (EXIF) metadata to determine whether the content of the target image can be generated by an imaging pipeline. Besides, Li \textit{et al.} \cite{li2018ictu} propose a deep fake face videos detection model based on LSTM and CNN, which utilizes the eye blinking signals of the target person in videos to determine whether the video is fabricated.

\subsection{Explanatory Misinformation Detection}
Exiting MID detection methods only give a final decision of whether a claim is false or not \cite{cao2018automatic}. Little information is revealed on why the decision is made. Finding evidences that support the decision would be beneficial in debunking the misinformation and preventing its further spreading. To this end, explanatory detection has become another trending research topic of MID.

In \cite{popat2017truth}, Popat \textit{et al.} put forward a probabilistic model to unite the content-aware and trend-aware evaluating algorithms for MID. Specifically, they model the mutual interactions among the event-related news articles for generating appropriate user-interpretable explanations, including linguistic features, stances and evidences, and the reliability of sources. Gad-Elrab \textit{et al.} \cite{gad2019exfakt} propose the \textit{ExFaKT} with the aim of providing human understandable explanations for candidate facts, which combines the semantic evidence from text and knowledge graphs. \textit{ExFaKT} uses Horn rules to rewrite the given fact as multiple easy-to-explain facts for further MID. \textit{ClaimVerif} \cite{zhi2017claimverif} is an explanatory claim credibility evaluation system, which takes the stance, viewpoint, source credibility and other factors of the given claim into account for providing evidences. Similarly, \textit{CredEye} \cite{popat2018credeye} is also an interpretable MID model which determines whether a given claim is fake by analyzing online articles related to it. The explanation is based on the language style, stance of these articles, and the credibility of their publishers.

\section{Crowd Intelligence-based Detection}
Existing MID studies mainly focus on the content of posts. However, as the posts are generated, interacted, and consumed by users, it will intake various \textit{human intelligence} (e.g., opinion, stance, questioning, evidence provision) in the creation, commenting, and reposting of posts, and the so-called \textit{crowd intelligence} \cite{li2017crowd, woolley2010evidence, guo2016mobile} is also aggregated at a collective manner during the dissemination process of a claim. As stated by \cite{castillo2011information}, a promising hypothesis is that there are some intrinsic signals in the social media environment which contribute to assessing the credibility of information. Ma \textit{et al.} \cite{ma2018rumor} also find that Twitter supports ``\textit{self-detect}'' of misinformation based on aggregated user opinions, conjectures and evidences. Though, how to leverage crowd intelligence in MID is still an open problem. In section 4, we attempt to address this problem by distilling and presenting several different forms of usage of crowd intelligence in MID systems.

\subsection{Crowd Intelligence in Misinformation}
In MID, crowd intelligence refers to aggregated cues or social signals from the wisdom of the social media users during the information generation and dissemination process. It can be characterized from the individual level and crowd level.

(1) \textbf{\textit{Individual level}}. It refers to the knowledge that can be leveraged from an individual.
\begin{itemize}
\item \textit{User profile}. User preferences, interests, cognition bias, demography, the number of followers.
\item \textit{User opinion and stance}. It involves opinions conveyed by users discussing the topic, for/against of the claim, enquires, questions, etc.
\item \textit{Source reliability}. It denotes the level of certainty of users who generate and propagate the information.
\end{itemize}

(2) \textbf{\textit{Crowd level}}. It refers to the knowledge that can be leveraged from a collective of social media users.
\begin{itemize}
\item \textit{Social contexts}. The social relations and interactions among source users and disseminators are helpful to understand the certainty of information.
\item \textit{Collective knowledge}. The collected evidences provided by the crowd are useful to infer the credibility of information.
\item \textit{Collective behaviors}. In many cases, though individual behaviors cannot well characterize the information credibility, the aggregated behaviors from a group of users often reveal more information. This may refer to crowd interaction patterns, behavior or opinion deviation from the majority \cite{kumar2018rev2}, conflicting viewpoints, and so on.
\end{itemize}

Having investigated existing MID studies, we distill four different manners of usage of crowd intelligence, as presented below.
\begin{itemize}
\item \textbf{\textit{Crowd learning models}}. It mainly uses feature engineering and representative learning to incorporate crowd intelligence in MID models.
\item \textbf{\textit{Crowd behavior modeling}}. It uses graph or probabilistic models to model crowd behaviors and interactions to infer the credibility of information.
\item \textbf{\textit{Crowd knowledge transferring}}. The learned MID models usually do not work well on new events. This manner tackles how to transfer crowd knowledge from existing events to new events.
\item \textbf{\textit{Hybrid human-machine models}}. Considering the complementary nature of human intelligence and machine intelligence, this manner concentrates on developing hybrid human-machine models for MID.
\end{itemize}

One common character of the prior three manners is that crowd intelligence is used in an implicit manner, without explicit human inputs. Specifically, crowd intelligence is represented as statistical human behavior patterns, used as features or parameters in the learning model. The last manner, however, is based on explicit human inputs, such as using crowdsourcing for data labeling. Thereafter, we describe related work about the prior three forms in Section 4.2, and the last form in Section 4.3.

\subsection{Implicit Crowd Intelligence Models}
In this section, we present the pioneering studies on the usage of implicit crowd intelligence for MID, particularly focusing on the first three manners depicted in Section 4.1, as summarized in Table~\ref{tab:CrowdIn}.

\begin{table}
 \caption{The Usage of Crowd Intelligence in MID}
 \label{tab:CrowdIn}
 \begin{tabular}{|m{4cm}<{\centering}|m{1.8cm}<{\centering}|m{2cm}<{\centering}|m{4cm}<{\centering}|}
 \hline
 \textbf{Usage manner} & \textbf{Work} & \textbf{Problem tackled} & \textbf{Usage of crowd intelligence} \\
 \hline
 \multirow{4}*{Crowd learning models} & Liu \textit{et al.} \cite{liu2015real} & Early detection & Features by collected opinion, belief, etc. \\
 \cline{2-4}
  & Zhao \textit{et al.} \cite{zhao2015enquiring} & Early detection & Features by crowd questions or enquires about the veracity. \\
 \cline{2-4}
  & Wu \textit{et al.} \cite{wu2018tracing} & General & Features by social relations and propagation network. \\
 \cline{2-4}
  & Rayana \textit{et al.} \cite{rayana2015collective} & General & Features by collective opinion clues and relational data. \\
 \hline
 \multirow{4}*{Crowd behavior modeling} & Hooi \textit{et al.} \cite{hooi2016birdnest} & General & Bayesian modeling, behavior deviation. \\
 \cline{2-4}
  & Kumar \textit{et al.} \cite{kumar2018rev2} & General & Bayesian modeling, behavior deviation. \\
 \cline{2-4}
  & Jin \textit{et al.} \cite{jin2016news} & Early detection & A credibility propagation network model that incorporates conflicting social viewpoints. \\
 \cline{2-4}
  & Ma \textit{et al.} \cite{ma2018rumor} & Early detection & Modeling reply structures and opinions by tree-structured recursive neural networks. \\
 \hline
 \multirow{3}*{Crowd knowledge transfer} & Wang \textit{et al.} \cite{wang2018eann} & Early detection \& Multimodal data fusion & The \textit{Event Adversarial Neural Network} model to derive event-invariant features. \\
 \cline{2-4}
  & Qian \textit{et al.} \cite{qian2018neural} & Early detection & A generative Conditional Variational Autoencoder to transfer user response knowledge. \\
 \cline{2-4}
  & Wu \textit{et al.} \cite{wu2017gleaning} & Early detection & A sparse representation model for shared feature learning. \\
 \hline
 \end{tabular}
\end{table}

\textbf{(1)} \textbf{\textit{Crowd learning models.}} In this model, crowd intelligence is represented as features to train classifiers for debunk misinformation. This has been proved useful for early detection of misinformation. For instance, Liu \textit{et al.} \cite{liu2015real} try to solve the problem of real-time rumor debunking using only crowd cues from Twitter data, including people's opinion, statistics of witness accounts, aggregated belief to the rumor, network propagation, and so on. Zhao \textit{et al.} \cite{zhao2015enquiring} observe that some people are willing to question or inquire the veracity of claims in Twitter before deciding whether to believe this message. Particularly, they find that the usage of enquiring minds facilitates early detection of low accuracy information.

Social relations and interactions are also widely-used crowd intelligence in MID feature learning. For instance, Wu \textit{et al.} \cite{wu2018tracing} assume that similar messages often conduce to similar information propagation traces. They propose a social media user embedding method to capture the features of social proximity and social network structures, atop which an LSTM model is utilized to classify the information propagation path and identify its veracity. In \cite{rayana2015collective}, Rayana \textit{et al.} apply collective opinion clues and relational data to detect misinformation.

It is also helpful to identify misinformation by leveraging the crowd intelligence that user behaviors of publishing misinformation diverges from those of posting genuine facts. Chen \textit{et al.} \cite{chen2018unsupervised} propose an unsupervised learning model which combines RNNs and Autoencoders to distinguish rumors from other authentic claims. In \cite{xie2012review}, Xie \textit{et al.} observe that the review spam attacks tend to be bursty and are strongly correlated with their rating patterns, which are distinct from the normal reviewers' behavior patterns. Therefore, they propose a review spam detection method based on their temporal behavior patterns.

\textbf{(2)} \textbf{\textit{Crowd behavior modeling.}} In this model, collective crowd behaviors, one type of crowd intelligence, are modeled as graphs or probabilistic models to infer information credibility. Hooi \textit{et al.} \cite{hooi2016birdnest} discover that fraudulent accounts which contribute to misinformation in rating systems often occur in short bursts of time, and have skewed distributions rating. The crowd wisdom is characterized by a Bayesian inference model, which can estimate how much a user's behaviors deviates from those of the related community and use it to infer information credibility. Similarly, Kumar \textit{et al.} \cite{kumar2018rev2} propose a Bayesian model that incorporate the aggregated crowd wisdom, such as the behavior properties of users, reliability of ratings, and goodness of products. By penalizing unusual behaviors, it can infer misinformation in rating platforms.

There are also studies that leverage aggregated crowd behavior modeling to facilitate early detection of misinformation. For example, Jin \textit{et al.} \cite{jin2016news} improve early detection of misinformation by mining conflicting viewpoints in microblogs. The supporting or opposing viewpoints are built into a credibility propagation network model for fake news detection. Ma \textit{et al.} \cite{ma2018rumor} assume that the repliers are inclined to enquiry who supports or denies the given rumor and express their desire for more evidence. They thus propose two tree-structured recursive neural networks (RvNN) for effective rumor representation learning and early detection, which can model the reply structures and learn to capture the aggregated rumor indicative signals.

\textbf{(3)} \textbf{\textit{Crowd knowledge transfer.}} Existing MID models still do not perform well on emerging and time-critical events. In other words, existing MID models usually capture abundant event-dependent features which are not common to other events. Therefore, it becomes a necessary to learn and transfer the shared knowledge learned from the existing crowdsourced data to new events. In \cite{wang2018eann}, Wang \textit{et al.} propose a detection model for identifying newly generated fake news using transferable features, named Event Adversarial Neural Network (\textit{EANN}), which comprises three parts, i.e., ``feature extractor'', ``event discriminator'', and ``fake news detector''. The \textit{EANN} uses the event discriminator to learn the event-independent sharing features, and reduces the influence of event-specific features on fake news detection.

Crowd knowledge transfer models also facilitate early detection of misinformation. For example, Qian \textit{et al.} \cite{qian2018neural} propose a generative Conditional Variational Autoencoder to capture user response patterns from the semantic information of historical users' comments on true and false news articles. In other words, crowd intelligence is leveraged to generate responses towards new articles to improve the detection capability of the model when user interaction data with articles are not available at the early stage of fake news propagation. Wu \textit{et al.} \cite{wu2017gleaning} also explore whether the detection of emerging rumors could be benefited by the knowledge acquired from historical crowdsourced data. They observe that similar rumors often lead to similar behavior patterns (e.g., curiosity, inquiry). A sparse representation model is built to select shared features and train event-independent classifiers.

\subsection{Hybrid Human-Machine Models}
MID is a challenging problem and merely automatic models cannot well adapt to various contexts and events. Human intelligence, however, can remedy this by leveraging their knowledge and experience. Hybrid human-machine models are thus developed to harness the complementary nature of human intelligence and machine intelligence for MID detection. Different from the implicit crowd intelligence-based models, the human intelligence used in such models are often layered on explicit human inputs. Such models often present sufficient explainability to facilitate human-computer collaboration, and with the aid of human-machine collaboration, the proposed models generally speed up the detection of misinformation.

\begin{table}
 \caption{The Usage of Hybrid Human-Machine Models}
 \label{tab:Hybrid}
 \begin{tabular}{|m{2cm}<{\centering}|m{3cm}<{\centering}|m{7cm}<{\centering}|}
 \hline
 \textbf{Work} & \textbf{Problem tackled} & \textbf{Usage of crowd intelligence} \\
 \hline
  Nguyen \textit{et al.} \cite{nguyen2018believe} & Explanatory detection & The mixed-initiative approach to blend human and machine intelligence. \\
 \hline
  Nguyen \textit{et al.} \cite{nguyen2018interpretable} & Explanatory detection & Incorporate explicit crowd intelligence in the probabilistic graphical model. \\
 \hline
  Vo \textit{et al.} \cite{vo2018rise} & Early detection & The machine recommends evidence URLs to human guardians to facilitate fact checking. \\
 \hline
  Kim \textit{et al.} \cite{kim2018leveraging} & Early detection & Using the marked temporal point processes to model crowd flagging procedure. \\
 \hline
  Tachiatschek \textit{et al.} \cite{tschiatschek2018fake} & Early detection & Using the Bayesian inference mode to incorporate crowd flagging behavior. \\
 \hline
  Lim \textit{et al.} \cite{lim2017ifact} & Explanatory detection \& Early detection & An interactive framework where machines collect evidences from Web search and human can give feedback to the evidences. \\
 \hline
  Bhattacharjee \textit{et al.} \cite{bhattacharjee2017active} & Early detection & An active learning model that introduces human-machine interaction to update the detection model. \\
 \hline
 \end{tabular}
\end{table}

There have been several human-machine models built for fact checking in MID, as summarized in Table ~\ref{tab:Hybrid}. For instance, Nguyen \textit{et al.} \cite{nguyen2018believe} consider that a system must be transparent in how it arrives at its prediction for users to trust the model. They propose a mixed-initiative approach that blends human knowledge and experience with AI for fact checking. In \cite{nguyen2018interpretable}, Nguyen \textit{et al.} present a hybrid human-machine approach based on the probabilistic graphical model, which integrates explicit human intelligence (by crowdsourcing) with the computing power to jointly model stance, veracity and crowdsourced labels. The approach is capable of generating interpretations to fact checking. Vo \textit{et al.} \cite{vo2018rise} present a fact-checking URL recommendation model with the purpose of stopping people from sharing misinformation. This model motivates guardians (users who tend to correct misinformation) to actively participate in fact-checking activities and spread the verified information to social networks.

Explicit human intelligence is also characterized and used in probabilistic models for fake news detection. In \cite{kim2018leveraging}, Kim \textit{et al.} propose \textit{CURB}, which leverages the marked temporal point processes to model crowd-powered flagging procedure for fake news. To significantly mitigate the propagation of misinformation with provable guarantees, \textit{CURB} can decide which claims to choose for fact checking and when to check misinformation. Tschiatschek \textit{et al.} \cite{tschiatschek2018fake} also present a Bayesian inference model that incorporates crowd flagging for detecting fake news. Lim \textit{et al.} \cite{lim2017ifact} present an interactive framework called \textit{iFACT} for assessing the credibility of new claims from tweets. It collects independent evidence from web search results and identifies the dependencies between historical claims and new claims. Users are allowed to provide explicit feedback on whether the web search results are relevant, support or against a claim. In \cite{bhattacharjee2017active}, Bhattacharjee \textit{et al.} propose a human-machine collaborative learning system for fast identification of fake news. An active learning approach is proposed, where an initial classifier is learnt on a limited amount of labelled data followed by an interactive approach to gradually update the model.

\section{Open Issues and Future Directions}
Though there have been increasingly considerable efforts to address the challenges in MID systems, there are still open issues to be studied in the future, as discussed below.

\textbf{(1)} \textbf{\textit{Model adaptivity/generality to new events}}. MID methods need to be capable of identifying unseen, newly coming events, since the existing data of the system may differ from that of emerging events. However, most existing approaches tend to extract event-specific features that hardly could be shared to newly-presented events \cite{zubiaga2018detection}. As stated by Tolosi \textit{et al.} \cite{tolosi2016analysis}, feature engineering-based methods can be hard to detect misinformation across different events as features change dramatically across events. Therefore, model generality or adaptivity is quite important for MID models to be applied to different events. Zubiaga \textit{et al.} \cite{zubiaga2017exploiting} state that the domain-dependent distribution of features could limit the generalization ability of the models. As the distribution of most characteristics direct to the event, the performance of the detection models will be affected. Though there have been some crowd knowledge transfer models \cite{wang2018eann, qian2018neural, wu2017gleaning} discussed in Section 4.2, there are much more to be investigated. Transfer learning models \cite{pan2009survey, guo2018citytransfer}, as successfully used in other domains, can be leveraged to design domain-adaptative MID models. The usage of GAN-based discriminators \cite{wang2018eann} is another promising way to generate generalized MID models with shared features.

Another interesting direction to be explored is that we should borrow knowledge from similar domains. For example, we can refer to web security, virus/spam detection methods \cite{zou2005monitoring, seneviratne2015early, chen2013cardinality}, which also suffer from similar issues such as early detection and model generalization.

\textbf{(2)} \textbf{\textit{Embracing of novel machine learning models}}. The misinformation detection process is by nature the learning of a classifier to identify the credibility of information. Recent years have witnessed the rapid and swift development of AI techniques. We have also found that many studies have built deep learning models \cite{ma2016detecting, chen2018call, ruchansky2017csi, yu2017convolutional, jin2017multimodal, nguyen2017early, liu2018early} to improve the performance of misinformation detection. Though, there are still more that can be explored. In the following we review several representative examples that leverage advanced machine learning techniques to MID.
\begin{itemize}
\item \textbf{\textit{Multi-task learning}}. Multi-task learning \cite{ma2018detect} is intended for improving the generalization performance of the model by using domain knowledge contained in related tasks. Existing multi-task learning methods seek to find commonalities among multiple tasks by modeling the task relevance, such as feature sharing, sub-space sharing, and parameter sharing, as a supplement to knowledge for promoting the learning effect of each task. For instance, Ma \textit{et al.} \cite{ma2018detect} consider that rumor detection is highly correlated with the stance classification task, and thus propose a neural multi-task learning framework for better detection. Under the mechanism of weight sharing, they present two RNN-based multi-task structures to jointly train these two tasks, which could extract ordinary as well as task-specific features for rumor representation. Inspired by this work, we can investigate the connection and collaboration between MID and other tasks and design multi-task learning based algorithms to improve MID performance.
\item \textbf{\textit{Semi-supervised models}}. Most existing MID works concentrate on supervised classification taking advantage of a great quantity of labeled data (e.g., fake or not) on the basis of manual feature extraction. However, in many cases we can only have a small number of labels. Semi-supervised models are often leveraged for dealing with the label sparsity issue. For example, Guacho \textit{et al.} \cite{guacho2018semi} propose a semi-supervised content-based MID method, which leverages the text embeddings based on tensor decomposition to capture the global and local information of the news article. After constructing the K-Nearest Neighbor (K-NN) graph of all the articles, they use a belief propagation algorithm to spread known article labels into the graph for obtaining the final credibility of the news.
\item \textbf{\textit{Hybrid learning models}}. The develop of hybrid learning models that integrate linear models with deep learning models have become a new research trend in AI, i.e., the combined usage of explicit features and latent features.  It is to leverage the complementary natures of both types of learning models. For example, \textit{Wide \& Deep} \cite{cheng2016wide} is a well-performed framework for recommender systems, where the \textit{Wide} part is used to integrate explicit features and the \textit{Deep} part is to learn the non-linear, latent features. There are also preliminary hybrid learning models in MID. Yang \textit{et al.} \cite{yang2018ti} propose the \textit{TI-CNN} model for fake news detection, which can be trained with textual and visual information corporately based on the fusion of explicit and implicit feature spaces. Though, hybrid learning models are still at its early stage, there are still much more to be studied towards this direction, such as the fusion of probabilistic graph models and deep learning models.
\end{itemize}

\textbf{(3)} \textbf{\textit{Explanatory detection models}}. Providing evidence or explanations to the learning results can increase user trust to the learning model. Though there have been few works on explanatory MID models, the introduction of explanations has been investigated in other related domains, such as recommender systems.

Explainable recommendation, which provides explanations about why an item is recommended, has attracted increasing attention in recent years \cite{zhang2018explainable}. It can help to improve users' acceptability, credibility, and satisfaction with recommender systems and enhance the systems' persuasiveness. For example, Chen \textit{et al.} \cite{chen2018visually} present a visually explainable recommendation method based on attentive neural networks to model user attention on images. Users can be informed of why an item is recommended by providing personalized and intuitive visual highlights. Catherine \textit{et al.} \cite{catherine2017explainable} study how to generate explainable recommendations with the support of external knowledge graphs, where a personalized PageRank procedure is proposed to rank items and knowledge graph entities together. In \cite{wang2018reinforcement}, Wang \textit{et al.} propose a model-agnostic explanatory recommendation system based on reinforcement learning (RL), which can flexibly control the presentation quality of the explanations. Above all, the methods used in such explainable recommendation systems can inspire us to design better explainable MID systems.

From a higher perspective, machine learning models have powered breakthroughs in diverse application areas (beyond recommender systems and MID). Despite the big success, we still lack understanding of their intrinsic behaviors, such as how a classifier arrives at a particular decision. This has result in the surging research direction of Interpretable Machine Learning (IML). IML gives machine learning models the ability to explain or to present in understandable terms to a human \cite{doshi2017towards, adadi2018peeking}. Du \textit{et al.} \cite{du2018techniques} define two types of interpretability: \textit{model-level interpretation} and \textit{prediction-level explanation}. Model-level interpretation can illuminate the inner working mechanisms of machine learning models and increase their transparency. Prediction-level explanation helps uncover the relations between a specific input and the model output. For MID, it is more focus on prediction-level explanation, which can illustrate how a decision can be arrived (using the elements such as source reliability, evidences, stances). A representative scheme of constructing prediction-level explainable models is employing the attention mechanism, which is widely utilized to explain predictions made by sequential models (e.g., RNNs). We should also study other approaches rooting from IML to enhance the explainability of MID systems.

\textbf{(4)} \textbf{\textit{Hybrid human-machine systems}}. As discussed in Section 4.3, the introduction of human intelligence in the detection of fake news has been proven a promising research direction. Broadly speaking, it belongs to the ``human computation'' paradigm, which aims to develop human-machine systems that interweave crowd and machine capabilities seamlessly to accomplish tasks that neither can do alone \cite{michelucci2016power, von2008human}. There have been several representative examples of human-machine systems. For example, \textit{reCAPTCHA} \cite{von2008recaptcha} is a  Captcha-like system to protect computer security, while at the same time it harnesses the combined efforts of individuals to the digitalization of books. \textit{Pandora} \cite{nushi2018towards} is a hybrid human-machine approach that cab explain failures in component-based machine learning systems.

MID has close relationship with another hot research topic in knowledge discovery, namely \textit{truth discovery}. Due to the ability to extract reliable information from conflicting multi-sourced information using human intelligence, truth discovery has become an increasingly significant research topic. For MID, we also have multiple posts about the same claim of an event, and the target is to identify the truth of the claim. Thereby, there are similarity between the two research problems and we can borrow knowledge from truth discovery systems to facilitate the MID. For example, to deduce the correct labels as much as possible from the noisy crowdsourced labels, Liu \textit{et al.} \cite{liu2017improving} propose an expert validation-assisted image label truth discovery method. In particular, it utilizes a semi-supervised learning algorithm which is able to maximize the influence of expert labels to lower expert efforts in human-computer collaboration. Zhang \textit{et al.} \cite{zhang2018texttruth} propose a probabilistic graph-based truth discovery model named ``\textit{TextTruth}'', which selects highly trustworthy answers to a given question by comprehensively learning the trustworthiness of key factors (a group of keywords) in each answer. The \textit{TextTruth} infers the credibility of the answer providers and the trustworthiness of every answer factor together in an unsupervised manner.

How to aggregate crowd wisdom is also important in MID systems. We can also learn from truth discovery systems. For example, Yin \textit{et al.} \cite{yin2017aggregating} present a model of crowd wisdoms aggregation in an unsupervised manner called Label-Aware Autoencoders (\textit{LAA}), which extracts the underlying features and patterns of multi-sourced labels and infers the trustworthy labels by a classifier and a reconstructor. To tackle the challenge that the same information source has different credibility on various topics, Ma \textit{et al.} \cite{ma2015faitcrowd} propose a crowdsourced data aggregation method named \textit{FaitCrowd}. The \textit{FaitCrowd} jointly learns the topic distribution of a given question, the specific topic-based knowledge of answer providers, and true answers by modeling the question content and the sources' response behaviors together on a probabilistic Bayesian model.

\textbf{(5)} \textbf{\textit{Propagation by social bots}}. Existing MID studies concentrate on the content and propagation patterns of posts. The characters of the ``accounts'' that publish and disseminate the posts, however, are not well investigated. Recently, there are several efforts paid to study the essential causes of misinformation spreading as rapidly as viruses. For example, Zhao \textit{et al.} \cite{shao2017spread} perform a detailed analysis of 14 million tweets posted on Twitter during the 2016 U.S. presidential election, 400 thousand of which contain misinformation. They observe that the ``social bots'' apparently facilitate the rapid diffusion of fake news. The social bot usually refers to a computer algorithm or software program that imitates human interaction behaviors (e.g., producing contents, following other accounts, reposting posts etc.) with purpose on social networks \cite{ferrara2016rise}. In addition, these malicious bot accounts are abnormally active in the very early stage of fake news propagation.

The above findings suggest that the suppression of social bots can be a promising way to mitigate the dissemination of misinformation. In the last few years, social bots have been active on social media platforms, such as Facebook, Twitter etc. There have also been studies on social bot behavior analysis and automatic detection. In \cite{ferrara2016rise}, Ferrara \textit{et al.} classify the existing detection approaches for social bots into four categories, includiing graph-based models, crowdsourcing, feature-based models, and hybrid models. Almaatoug \textit{et al.} \cite{almaatouq2016if} conduct an authoritative analysis of the behaviors of the spam accounts in social networks. They design a detection method that incorporate the features such as content attributes, social interactions, and profile properties. Similarly, Minnich \textit{et al.} \cite{minnich2017botwalk} propose the \textit{BotWalk} bot detection method, where they characterize accounts with various behavioral features, such as metadata, content, temporal information, and network interaction. Cresci \textit{et al.} \cite{cresci2017social} conduct a penetrating analysis of the collective behaviors of social bots and present the \textit{Social Fingerprinting} technique for spambot detection. In particular, they exploit the \textit{digital DNA} technique for characterizing the collective behaviors of all the accounts, and a DNA-inspired method is then proposed to characterize genuine accounts and spambots. Cresci \textit{et al.} \cite{cresci2017paradigm} also leverage the characteristics of group accounts to detect spambots.

\textbf{(6)} \textbf{\textit{Adversarial attack and defense in misinformation detection models}}. As discussed in Section 2.4, the prevalence of deep learning-based MID models contributes to effective improvement of recognition performance. However, Szegedy \textit{et al.} \cite{szegedy2013intriguing} have proven that the trained neural networks may fail to respond to adversarial attacks, which means that adding some small perturbations to the input vectors could make models get wrong results \cite{akhtar2018threat}. Existing MID studies rarely highlight the robustness of detection models which can be deceived by adversarial attacks.

Although there have been few studies on adversarial attack and defense in MID models, the related work about other tasks \cite{goodfellow2014explaining, kuleshov2018adversarial} has been investigated. For instance, Dai \textit{et al.} \cite{dai2018adversarial} present an adversarial attack method for graph data based on reinforcement learning which learns the optimal attack policy by increasing or decreasing the amount of edges in the graph. With the purpose of generating universal adversarial perturbations for text, Behjati \textit{et al.} \cite{behjati2019universal} propose a gradient projection-based attack method. Jia \textit{et al.} \cite{jia2017adversarial} attack question and answer systems by adding sentences or phrases to questions that do not cause difficulty to the human understanding.

The above attack studies can provide guidance for the research of adversarial attack defense in MID models. Zhou \textit{et al.} \cite{zhou2019fake} further divide adversarial attacks on MID models into \textit{fact distortion}, \textit{subject-object exchange}, and \textit{cause confounding}, and propose a crowdsourced knowledge graph to collect timely facts about news events for resisting adversarial attacks. Qiu \textit{et al.} \cite{qiu2019review} classify defensive methods into three categories, including \textit{modifying data} (e.g. adversarial training, gradient hiding), \textit{modifying models} (e.g. regularization, defensive distillation), and \textit{using auxiliary tools} (e.g. \textit{Defense-GAN} \cite{samangouei2018defense}). Though some studies have already been done in this field, there are still more efforts to be conducted on adversarial attack and defense for MID.

\section{Conclusion}
We have made a systematic review of the research trends of misinformation detection. Having given a brief review of the literature of MID, we present several new research challenges and techniques of it, including early detection, detection by multimodal data fusion, and explanatory detection. We further investigate the usage of crowd intelligence in MID, including crowd intelligence-based MID models and hybrid human-machine MID models. Though there has been a big research progress in MID, it is still at the early stage and there are numerous open research issues and promising research directions to be studied, such as model adaptivity/generality to new events, embracing novel machine learning models, explanatory detection models, and so on.

%%
%% The acknowledgments section is defined using the "acks" environment
%% (and NOT an unnumbered section). This ensures the proper
%% identification of the section in the article metadata, and the
%% consistent spelling of the heading.
\begin{acks}
This work was partially supported by the National Key R\&D Program of China (2017YFB1001803), and the National Natural Science Foundation of China (No. 61772428, 61725205).
\end{acks}

%%
%% The next two lines define the bibliography style to be used, and
%% the bibliography file.
\bibliographystyle{ACM-Reference-Format}
\bibliography{Misinformation_Ref}

\end{document}